\documentclass[a4paper]{cas-dc}
\usepackage[numbers,sort&compress]{natbib}
\usepackage{braket}
\usepackage{multicol}
\usepackage{subfigure}
\usepackage{ulem}

\definecolor{darkpastelgreen}{rgb}{0.01, 0.75, 0.24}
\definecolor{electricindigo}{rgb}{0.44, 0.0, 1.0}
\definecolor{palatinateblue}{rgb}{0.15, 0.23, 0.89}
\definecolor{carminered}{rgb}{1.0, 0.0, 0.22}
\hypersetup{linkcolor=palatinateblue, citecolor=electricindigo, urlcolor=carminered}

\begin{document}
\shorttitle{Dominance of gluonic scale anomaly in confining pressure inside nucleon and D-term}
\shortauthors{D. Fujii, M. Kawaguchi and M. Tanaka}  
\title[mode = title]{Dominance of gluonic scale anomaly in confining pressure inside nucleon and D-term}  

\author[1,2]{Daisuke Fujii}[type=editor, orcid=0000-0002-6298-9278]
\nonumnote{$\dag$ E-mail: daisuke@rcnp.osaka-u.ac.jp (corresponding author)}

\author[3]{Mamiya Kawaguchi}[orcid=0000-0002-3103-1315
]
\nonumnote{$\ddag$ E-mail: mamiya@aust.edu.cn
(corresponding author)}

\author[4]{Mitsuru Tanaka}[orcid=0009-0000-5341-562X]
\nonumnote{$\ddag$ E-mail: tanaka@hken.phys.nagoya-u.ac.jp (corresponding author)}

\address[1]{Advanced Science Research Center, Japan Atomic Energy Agency (JAEA), Tokai, 319-1195, Japan}
\address[2]{Research Center for Nuclear Physics, Osaka University, Ibaraki 567-0048, Japan}
\address[3]{Center for Fundamental Physics, School of Mechanics and Physics, Anhui University of Science and Technology, Huainan, 232001, People’s Republic of China}
\address[4]{Department of Physics, Nagoya University, Nagoya 464-8602, Japan}

\begin{abstract}
We explore the confining pressure inside the nucleon and the related gravitational form factor referred to as the D-term, using the skyrmion approach based on the scale-invariant chiral perturbation theory, where the skyrmion is described as the nucleon and a scalar meson couples to the scale anomaly through the low energy theorem. 
Within this model framework, the current quark mass and gluonic quantum contributions to the scale anomaly can be described by the pion and scalar meson masses, respectively, through matching with the underlying QCD.
By considering the decomposition of the energy momentum tensor of nucleon, we examine the role of the scale anomaly contributions in the pressure inside the nucleon. 
As a result, the gluonic scale anomaly is found to dominate the confining pressure. 
Compared to the result based on the conventional chiral perturbation theory in the chiral limit, our result for the total pressure is capable of qualitatively improving the alignment with lattice QCD observations. Moreover, the pressure from the gluonic scale anomaly is widely distributed in position space, leading to its substantial contribution to the D-term.
\end{abstract}


\maketitle

\section{Introduction}
Uncovering the mechanisms by which quarks and gluons are confined inside hadrons is one of the most fundamental challenges in quantum chromodynamics (QCD). Critical insight into this issue is anticipated to arise from investigating the role of non-perturbative properties of QCD, such as the quark-antiquark pair condensate (chiral condensate) and the gluon condensate, as well as symmetry breaking.

In 2018, the stress distribution (pressure and shear force) inside the proton was experimentally measured \cite{Burkert:2018bqq,Burkert:2021ith,Duran:2022xag}, which has led to a rapid progress in theoretical studies \cite{Polyakov:2018exb,Shanahan:2018nnv,Lorce:2018egm,Anikin:2019kwi,Avelino:2019esh,Yanagihara:2019foh,Hatta:2019lxo,Freese:2019eww,Azizi:2019ytx,Mamo:2019mka,Neubelt:2019sou,Alharazin:2020yjv,Varma:2020crx,Kim:2020nug,Chakrabarti:2020kdc,Yanagihara:2020tvs,Kim:2020lrs,Tong:2021ctu,Freese:2021czn,Panteleeva:2021iip,Hatta:2021can,Mamo:2021krl,Freese:2021qtb,Gegelia:2021wnj,Kim:2021jjf,Owa:2021hnj,Pefkou:2021fni,Lorce:2021xku,Ji:2021mfb,More:2021stk,Mamo:2022eui,Lorce:2022cle,Fujita:2022jus,Choudhary:2022den,Kim:2022wkc,Alharazin:2022wjj,Won:2022cyy,Tanaka:2022wzy,Ito:2023oby,Lorce:2023zzg,Amor-Quiroz:2023rke,Guo:2023pqw,Won:2023ial,Guo:2023qgu,Czarnecki:2023yqd,Won:2023zmf,Hackett:2023rif,Hatta:2023fqc,Liu:2023cse,Cao:2024zlf,Liu:2024rdm,Yao:2024ixu,Goharipour:2025lep,Dehghan:2025ncw,Goharipour:2025yxm,Broniowski:2025ctl,Ghim:2025gqo,Hatta:2025vhs,Hatta:2025ryj}. 
This stress distribution is extracted from the form factors that characterize the matrix elements of the energy-momentum tensor (EMT) for hadrons, which are referred to as the gravitational form factors (GFFs) (for reviews, see Refs.~\cite{Polyakov:2018zvc,Burkert:2023wzr}). 
This mechanical distribution represents the forces that confine quarks and gluons inside hadrons, providing a pathway to address the above issue. 
However, it remains unclear which non-perturbative properties crucially contribute to the confining pressure and the related GFFs.

To address this question, we focus on the scale anomaly in this letter, which is one of the non-perturbative properties of QCD, and explore its role in the stress distribution inside the nucleon and the related GFFs.
Indeed, the scale anomaly is directly associated with the EMT $\Theta_{\mu\nu}$, which is expressed as 
\begin{align}
    {\Theta^\mu}_\mu=
    \partial_\mu J^\mu_D=(1+\gamma_m)\sum_fm_f\bar q_fq_f+\frac{\beta(g_s)}{2g_s}{\rm Tr}\left(G_{\mu\nu}G^{\mu\nu}\right), \label{traceanomaly}
\end{align}
where
$J^\mu_D$ denotes the dilatation current,
$G_{\mu\nu}$ is the field strength of $SU(N_c)$ gluon fields, $m_f$ are the current quark masses and $q, \bar q$ are quark and anti-quark fields with flavor indices $f$.
This conservation law clearly shows that the 
scale invariance is broken by the current quark mass with 
the anomalous dimension $\gamma_m$ and by the gluonic quantum correction characterized by the beta function $\beta(g_s)$ with the strong coupling $g_s$.
Indeed, these anomalous contributions to the scale invariance significantly influence the pressure inside the nucleon and the GFFs, as will be discussed later. 

In general, when investigating the role of the non-perturbative properties of QCD, 
the effective model is a robust approach and provides a rigorous framework that is built upon the fundamental properties of the underlying QCD theory. 
In this letter, we employ the Skyrmion model approach~\cite{Skyrme:1962vh,Adkins:1983ya}, 
which successfully describes the nucleon from the viewpoint of the large $N_c$ expansion~\cite{tHooft:1973alw,Witten:1979kh}. 
Furthermore, 
our model is 
based on the scale-invariant chiral perturbation theory (sChPT)~\cite{Lanik:1984fc,Ellis:1984jv,Leung:1989hw,Campbell:1990ak,Donoghue:1991qv,Brown:1991kk,Song:1997kx,Lee:2003eg,Park:2003sd,Park:2008zg,Li:2018gng,Crewther:2013vea,Matsuzaki:2013eva,Li:2016uzn,Kasai:2016ifi,Hansen:2016fri,Appelquist:2017wcg,Appelquist:2017vyy,Cata:2019edh,Appelquist:2019lgk,Brown:2019ipr}, which reflects the scale anomaly of the underlying QCD theory in Eq.~\eqref{traceanomaly}. 
Note that the conventional chiral perturbation theory (ChPT) is constructed to reproduce the QCD low-energy theorem of the chiral symmetry, i.e., the partially conserved axial vector current (PCAC) relation~\cite{Gasser:1983yg,Gasser:1984gg}, but does not take into account that of the scale symmetry. 
To capture the scale anomaly through the low-energy theorem, the ChPT is extended to the sChPT by introducing a scalar meson, which couples to 
$J^\mu_D$. 
Indeed, many previous studies have explored how the pressure is described in terms of quark and gluon degrees of freedom~\cite{Shanahan:2018nnv,Lorce:2018egm,Neubelt:2019sou,Guo:2023pqw,Won:2023ial,Guo:2023qgu,Won:2023zmf,Hackett:2023rif,Yao:2024ixu}. 
However, no detailed research has specifically focused on the non-perturbative nature of the scale anomaly 
in relation to the pressure stabilized in the nucleon.
In light of this fact, we address this issue using the skyrmion approach in sChPT to clarify the role of the scale anomaly in the pressure
and GFFs.

What we have done and found is summarized as follows.
First, it is found that the confining pressure inside the nucleon arises from the scale anomaly, dominantly driven by the gluonic contribution. 
Second, compared to the conventional ChPT in the chiral limit~\cite{Cebulla:2007ei,GarciaMartin-Caro:2023toa}, in which the scale symmetry is not appropriately reflected, 
the pressure evaluated by using the sChPT becomes more aligned with the lattice QCD observation~\cite{Shanahan:2018nnv}. 
Third, 
we find that the contribution of the gluonic scale anomaly dominates the D-term (defined in a forward limit $D=D(t=0)$ of the GFF $D(t)$). Although the D-term 
is referred to as the {\it "last unknown global property"}~\cite{Polyakov:2018zvc} of the nucleon, 
given that our pressure result aligns with the lattice observation~\cite{Shanahan:2018nnv}, the model provides a reliable prediction of the D-term, which is consistent with the recent lattice observation within its systematic errors~\cite{Hackett:2023rif}.

\section{Pressure and D-term decomposed in terms of the scale anomaly}

To facilitate our discussion,
we first provide a brief review of how the pressure is defined inside the nucleon.
We then decompose the pressure in terms of the scale anomaly and connect the decomposed pressures to the GFFs.

We start with the matrix elements of the EMT sandwiched between nucleon states:
\begin{align}
    &\Braket{p',s'|\Theta_{\mu\nu}(x)|p,s} \notag \\
    &=\bar{u}'\Big[A(t)\frac{P_\mu P_\nu}{M_N}+J(t)\frac{i\left(P_\mu\sigma_{\nu\rho}+P_\nu\sigma_{\mu\rho}\right)\Delta^\rho}{2M_N} \notag \\
    &\hspace{10mm}+D(t)\frac{\Delta_\mu\Delta_\nu-g_{\mu\nu}\Delta^2}{4M_N}\Big]ue^{i(p'-p)x}, \label{GFFs}
\end{align}
where $A(t), \ J(t), \ D(t)$ are the GFFs; $\sigma_{\mu\nu}$ is defined as $\sigma_{\mu\nu}=\frac{i}{2}[\gamma^\mu,\gamma^\nu]$ with the Dirac matrix $\gamma_\mu$; the momentum $P^\mu$ and $\Delta^\mu$ are defined by $P^\mu=(p^\mu+p'^\mu)/2$ and $\Delta^\mu=p'^\mu-p^\mu$, and $t=\Delta^2$;
$M_N$ is a nucleon mass; 
$s,s'=\pm1/2$ represents the spin state of a hadron;
$u(p,s)$ is the Dirac spinor. 
The stress distribution inside the nucleon is obtained from the Fourier transform of the matrix elements of the EMT in the Breit frame where $P_\mu=(E,0,0,0), \Delta_\mu=(0,\vec{\Delta}), t=-\vec{\Delta}^2$. 
This formulation represents the static EMT of the nucleon:
\begin{align}
\Theta^{\rm static}_{\mu\nu}(\vec{r},\vec{s})=\int\frac{d^3\vec{\Delta}}{(2\pi)^3}e^{-i\vec{r}\cdot\vec{\Delta}}\frac{\Braket{p',s'|\Theta_{\mu\nu}(x=0)|p,s}}{\bar u(p')u(p)}, \label{staticEMT}
\end{align}
where $\vec{s}$ is the polarization vector of the states $\ket{p,s}$ and $\ket{p',s'}$, and the normalization is given by $\bar u(p')u(p)=2E$ 
with the energy $E=\sqrt{M_N^2+\vec{\Delta}^2/4}$
and 
$\braket{p'|p}=2p^0(2\pi)^3\delta^{(3)}(\vec{p}'-\vec{p})$.
From the conservation law of EMT ($\partial^\mu\Theta_{\mu\nu}=0$), the static EMT satisfies $\partial^i\Theta^{\rm static}_{ij}=0$. 
Using this static EMT, the pressure inside the nucleon can be defined as $p(r) =\delta^{ij}\Theta^{\rm static}_{ij}/3$ 
for a spherically symmetric system.
Furthermore, from the conservation law of static EMT, the pressure satisfies the von Laue condition $\int^\infty_0drr^2p(r)=0$, 
which represents the balance relation of pressure distribution in the stable system.

To clearly extract the contribution of the scale anomaly, we consider the following decomposition of the EMT, separating the scale anomaly,
\begin{align}
    &\Theta^{\rm static}_{\mu\nu}(\vec{r},\vec{s})=\bar{\Theta}^{\rm static}_{\mu\nu}+\hat{\Theta}^{\rm static}_{\mu\nu} \label{staticEMTdecomp} \\
    &\bar{\Theta}^{\rm static}_{\mu\nu}(\vec{r},\vec{s})=\int\frac{d^3\vec{\Delta}}{(2\pi)^3}e^{-i\vec{r}\cdot\vec{\Delta}}\frac{\Braket{p',s'|\bar{\Theta}_{\mu\nu}(0)|p,s}}{\bar u(p')u(p)}\\
    &\hat{\Theta}^{\rm static}_{\mu\nu}(\vec{r},\vec{s})=\int\frac{d^3\vec{\Delta}}{(2\pi)^3}e^{-i\vec{r}\cdot\vec{\Delta}}\frac{\Braket{p',s'|\hat{\Theta}_{\mu\nu}(0)|p,s}}{\bar u(p')u(p)}.
\end{align}
$\bar{\Theta}_{\mu\nu}$ corresponds to the traceless part of the EMT, $\bar{\Theta}_{\mu\nu}=\Theta_{\mu\nu}-\frac{1}{4}g_{\mu\nu}{\Theta^\rho}_\rho$, 
which stems from the dynamical energy of quarks and gluons and $\hat{\Theta}_{\mu\nu}$ represents the trace part, defined as $\hat{\Theta}_{\mu\nu}=\frac{1}{4}g_{\mu\nu}{\Theta^{\rho}}_\rho$, which corresponds to the scale anomaly. 
In the following, we consider the case $s=s'$.

From the static EMT in Eq.~(\ref{staticEMTdecomp}), the pressure $p(r)$ is decomposed as 
\begin{align}
    &p(r)=
    \bar{p}(r)+\hat{p}(r) \label{pdecomp} \\
    &\bar{p}(r)=\frac{1}{3}\delta^{ij}\bar{\Theta}^{\rm static}_{ij} \\
    &\hat{p}(r)=\frac{1}{3}\delta^{ij}\hat{\Theta}^{\rm static}_{ij}=-\frac{1}{4}\left({\Theta^\rho}_\rho\right)^{\rm static}.
\end{align}
On the other hand, the energy density of the nucleon can also be obtained as ${\epsilon}(r)={\Theta}^{\rm static}_{00}$.
In the same way as the pressure,
this energy density inside the nucleon can be decomposed into two parts:
${\epsilon}(r) =
\bar \epsilon
+ 
\hat \epsilon
$
with $\bar \epsilon = \bar{\Theta}^{\rm static}_{00}$ and 
$\hat \epsilon = \hat{\Theta}^{\rm static}_{00}$. 
When performing the spatial integral of the energy density, it yields the mass of the nucleon, which can also be decomposed into two parts: one originating from the dynamical energy of the quarks and gluons, $\bar M_N=4\pi\int^\infty_0drr^2\bar\epsilon(r)$, and the other from the scale anomaly, $\hat M_N=4\pi\int^\infty_0drr^2\hat\epsilon(r)$~\cite{Ji:1994av}.
Using Eq.~\eqref{GFFs} with $p=p', \ s=s'$ and $A(0)=1$,
the traceless part and trace part of the nucleon mass are given by~\cite{Ji:1994av}:
\begin{align}
\bar M_N=\frac{3}{4}M_N, \ \ \ \hat M_N=\frac{1}{4}M_N.
\label{1/4M}
\end{align}
The contribution of the scale anomaly accounts for $25\%$ of the nucleon mass.

Thanks to the decomposition of the EMT,
one can gain insight into the details of the pressure~\cite{Lorce:2017xzd,Lorce:2021xku}. Performing the spatial integration of $\hat p$ gives the following result,
\begin{align}
    4\pi\int^\infty_0drr^2\hat p(r)=-\frac{1}{4}M_N. \label{hatp}
\end{align}
Furthermore, applying the von Laue condition, 
the spatial integration of $\bar p$ is evaluated as 
\begin{align}
    4\pi\int^\infty_0drr^2\bar p(r)=\frac{1}{4}M_N. \label{barp}
\end{align} 
The signs of these two equations provide the information about the stability of the nucleon:
the dynamical energy of quarks and gluons, described by the spatial integration of $\bar p(r)$, provides a repulsive effect while  
the contribution of
the scale anomaly, captured by the integration of $\hat p(r)$, has a confining effect.
Namely, $\bar p(r)$ and $\hat p(r)$ are balanced inside the nucleon from a global perspective, as obtained by performing the spatial integral, implying that the scale anomaly plays an important role in providing the confining pressure necessary for stability. However, to understand the detailed behavior of the pressure from a local perspective, as seen in the $r$-dependence of $\bar p$ and $\hat p$, it is necessary to either solve QCD directly or use effective model approaches. 

In addition,
the stability of the pressure inside the nucleon is reflected in the GFFs, particularly the D-term. By using the decomposed pressure, the D-term can be expressed as~\cite{Ji:2021mtz}
\begin{align}
    D=M_N\int d^3rr^2p(r)=\frac{1}{4}M_N^2\left(\braket{r^2}_t-\braket{r^2}_s\right), \label{D}
\end{align}
where the tensor mean square radius $\braket{r^2}_t$ and scalar mean square radius $\braket{r^2}_s$, are given by $\braket{r^2}_t=\int d^3rr^2\bar p(r)/M_N$ and $\braket{r^2}_s=-\int d^3r^2\hat p(r)/M_N$. 
Indeed, as discussed in Ref.~\cite{Perevalova:2016dln}, 
the D-term is estimated to be negative in the stable matter, which results in the scalar radius being larger than the tensor radius. 
The negativity of the D-term is essential for ensuring stability, again showing that the scale anomaly plays an important role in the stability of the nucleon.
However, the details of the scale anomaly contribution remain unclear: 
which of the two ingredients in Eq.~(\ref{traceanomaly}), the current quark mass part or the gluonic quantum correction, plays a dominant role?

\section{Static EMT of skyrmion based on sChPT}

To obtain the local information of pressure and clarify which ingredient of the scale symmetry breaking dominates the negativity of the D-term,  
we employ the effective model approach.
As mentioned in the introduction, this study uses the sChPT to describe the skyrmion as a nucleon.
The conventional ChPT is constructed based on spontaneous chiral symmetry breaking, which is formulated in terms of the derivative expansion~\cite{Gasser:1983yg,Gasser:1984gg}.
The leading order of the ChPT Lagrangian is given by
$\mathcal{L}^{[2]}_{\mathcal{O}\left(p^2\right)}=\frac{f_\pi^2}{4}g^{\mu\nu}{\rm Tr}\left(\partial_\mu U^\dagger\partial_\nu U\right)$ and $\mathcal{L}^{[0]}_{\mathcal{O}\left(m_\pi^2\right)}=\frac{1}{4}f_\pi ^2m_\pi^2{\rm Tr}\left(U+U^\dagger\right)$, 
where $[d]$ is the scale dimension of Lagrangian density; the chiral field $U$, given by
$U=\exp\left(i\pi^a\tau^a/f_\pi\right)$ with the pion field $\pi^a$ and the Pauli matrix $\tau^a$ ($a=1,2,3$),
is the building block of the ChPT Lagrangian
and represents the scale dimension of $0$;
$f_\pi$ and $m_\pi$ are the decay constant of pion and the pion mass, respectively.
Indeed, effective models should satisfy the low energy theorem required by the underlying QCD.
This theorem is described by an overlap amplitude between pions and the broken current associated with the $SU(2)$ axial symmetry:
\begin{eqnarray}
\Braket{0|\partial_\mu J^{a\mu}_{5}(x)|\pi^b(p)} &=&- f_\pi m_\pi^2 
e^{-ip\cdot x}
\delta^{ab}
\label{chiralbreaking},
\end{eqnarray}
with $J^{a\mu}_{5}$ being axial vector current.
The ChPT Lagrangian, which includes the explicit symmetry breaking in the form of the pion mass term, surely provides this theorem. Hence, the $SU(2)$ axial symmetry of the underlying QCD theory is appropriately reflected in the ChPT Lagrangian. However, the information related to the scale symmetry of the underlying QCD, particularly the scale anomaly in Eq.~(\ref{traceanomaly}), is not included in ChPT.

To incorporate the scale symmetry-breaking into the effective model, we employ the sChPT, which is constructed to respect a different low energy theorem associated with the scale symmetry. 
This theorem is also described by an overlap amplitude where the lightest isoscalar particle is assumed to couple to the dilatation current, known as the partially conserved dilatation current (PCDC) relation ,
\begin{eqnarray}
\Braket{0|\partial_\mu J_D^\mu(x)|\phi(p)}
&=&
- f_\phi m_\phi^2 
e^{-ip\cdot x}.
\end{eqnarray}
where $\phi$ is regarded as a lightest scalar meson;
$f_\phi$ and $m_\phi$ are the decay constant and mass of the scalar meson.
To satisfy the PCDC relation as well as the PCAC relation, the sChPT Lagrangian is described by~\cite{Lanik:1984fc,Ellis:1984jv,Campbell:1990ak,Donoghue:1991qv,Brown:1991kk,Song:1997kx,Lee:2003eg,Park:2003sd,Park:2008zg,Li:2018gng,Crewther:2013vea,Li:2016uzn}
\begin{align}
    &\mathcal{L}_{\rm sChPT}=\left(\frac{\chi}{f_\phi}\right)^2\mathcal{L}^{[2]}_{\mathcal{O}\left(p^2\right)}+\left(\frac{\chi}{f_\phi}\right)^{3-\gamma_m}\mathcal{L}^{[0]}_{\mathcal{O}\left(m_\pi^2\right)} \notag \\
    &\hspace{16mm}+\mathcal{L}^\chi_{\rm kin}+\mathcal{L}^\chi_{\rm pot} \label{sChPTL} \\
    &\mathcal{L}^\chi_{\rm kin}=\frac{1}{2}g^{\mu\nu}\partial_\mu\chi\partial_\nu\chi \\
    &\mathcal{L}^\chi_{\rm pot}=
    -\frac{1}{4}m_{\phi0}^2f_\phi^2\left(\frac{\chi}{f_\phi}\right)^4\left[\ln\frac{\chi}{f_\phi}-\frac{1}{4}\right]
\end{align}
where
$\chi$ is the conformal compensator with the scale dimension of $1$, nonlinearly parametrized by $\phi$, such that $\chi=f_\phi e^{\phi/f_\phi}$;
$m_{\phi0}$ represents the scalar meson mass defined in the chiral limit.
Furthermore, in this model, 
the vacuum expectation value of the scale anomaly is evaluated as 
\begin{align}
    \Braket{0|\partial^\mu J^D_\mu(x)|0}=-\left(1+\gamma_m\right)f_\pi^2m_\pi^2-\frac{f_\phi^2m_{\phi0}^2}{4}. \label{sanomaly_vac}
\end{align}
Considering the Gell-Mann-Oakes-Renner relation, $f_\pi^2m_\pi^2=-m_f\braket{0|\bar qq|0}$, the first term in Eq.~(\ref{sanomaly_vac}) comes to correspond to the current quark mass part of the scale anomaly in Eq.~(\ref{traceanomaly}).
On the other hand, the remaining second term in Eq.~(\ref{sanomaly_vac}) would be matched with the gluonic part in Eq.~(\ref{traceanomaly}).
Namely, within the sChPT framework, the chiral-limit mass of the scalar meson $m_{\phi0}$ with the decay constant mimics the gluonic quantum correction in the scale anomaly of the underlying QCD: $f_\phi^2 m_{\phi0}^2/4 =- \langle0| (\beta(g_s)/2g_s){\rm Tr}\left(G_{\mu\nu}G^{\mu\nu}\right)  |0 \rangle $. 
Note that, from the sChPT Lagrangian, the physical scalar meson mass $m_\phi$ is given by 
\begin{align}
    &m_\phi^2 = m_{\phi0}^2 -(3-\gamma_m)^2\frac{f_\pi^2m_\pi^2}{f_\phi^2}
    \nonumber\\
    &=- \frac{4}{f_\phi^2}\langle0| \frac{\beta(g_s)}{2g_s}{\rm Tr}\left(G_{\mu\nu}G^{\mu\nu}\right)  |0 \rangle
    + (3-\gamma_m)^2\frac{m_f \langle0| \bar qq|0\rangle}{f_\phi^2}.
\end{align}
This shows that, in the chiral limit, the scalar meson mass is solely provided by the gluonic quantum correction. When considering QCD with the massive current quark, the quark condensate associated with the current quark mass part also contributes to the scalar meson mass.

In the effective model, 
to connect the energy momentum tensor with the divergence of the dilatation current, the improved energy momentum tensor is defined: $\Theta_{\mu\nu}=T_{\mu\nu}+\theta_{\mu\nu}$
where $T_{\mu\nu}$ is the conserved Noether current associated with spacetime translations and $\theta_{\mu\nu}$ is the improvement term, 
such that ${\Theta^\mu}_\mu=\partial_\mu J^\mu_D$. Note that in the effective model, $J^\mu_D$ is redefined by using $\theta_{\mu\nu}$.
At the operator level of the sChPT, the traceless part of the EMT is evaluated as
\begin{align}
    &\bar{\Theta}_{\mu\nu}=\frac{f_\pi^2}{2}\left(\frac{\chi}{f_\chi}\right)^2{\rm Tr}\left(\partial_\mu U^\dagger\partial_\nu U\right)+\frac{2}{3}\partial_\mu\chi\partial_\nu\chi-\frac{1}{3}\chi\partial_\mu\partial_\nu\chi \notag \\
    &\hspace{12mm}-g_{\mu\nu}\left\{\frac{f_\pi^2}{8}\left(\frac{\chi}{f_\chi}\right)^2{\rm Tr}\left(\partial_\rho U^\dagger\partial^\rho U\right)\right. \notag \\
    &\left.\hspace{24mm}+\frac{1}{6}\partial_\rho\chi\partial^\rho\chi-\frac{1}{12}\chi\partial_\rho\partial^\rho\chi\right\}. \label{barEMTsChPT}
\end{align}
Within the sChPT framework, the scale anomaly from the current quark mass (gluonic quantum correction) is recast as the contribution of the pion (scalar meson) mass, which is incorporated into the trace part: 
\begin{align}
    &\hat\Theta_{\mu\nu}=\hat{\Theta}_{\mu\nu}^q+\hat{\Theta}_{\mu\nu}^g \label{hatEMTsChPT} \\
    &\hat{\Theta}_{\mu\nu}^q=-\frac{1}{4}g_{\mu\nu}
    (1+\gamma_m)
    \frac{f_\pi^2m_\pi^2}{4}\left(\frac{\chi}{f_\chi}\right)^{3-\gamma_m}{\rm Tr}\left(U+U^\dagger\right), \label{hatEMTsChPTq} \\
    &
    \hat{\Theta}_{\mu\nu}^g=-\frac{1}{4}g_{\mu\nu}\frac{m_{\phi0}^2f_\phi^2}{4}\left(\frac{\chi}{f_\phi}\right)^4.
    \label{hatEMTsChPTg}
\end{align}
Since the contribution of the pion mass can be distinct from that of the scalar meson mass, the traceless part is further decomposed into the quark and gluon parts, as shown in Eq.~(\ref{hatEMTsChPT}).

To describe the nucleon, we adopt the skyrmion approach~\cite{Skyrme:1962vh,Adkins:1983ya}, where the nucleon is represented as a static soliton, called skyrmion.
This skyrmion is characterized by the topological charge related to the baryon number, and is a spatially extended object that possesses a finite size.
To stabilize the static soliton and ensure it has finite energy in the effective model, a higher-derivative term, known as the Skyrme term, is introduced:
$\mathcal{L}^{[4]}_{\mathcal{O}(p^4)}=\frac{1}{32e^2}{\rm Tr}([U^\dagger\partial_\mu U,U^\dagger\partial_\nu U]^2)$ with the Skyrme parameter $e$.
In the skyrmion approach, to obtain the soliton solution, it is assumed that the chiral field takes the form of the hedgehog ansatz, $U(\vec{x})=\exp \big(i\vec{\tau}\cdot\vec{\hat{r}}F(r)\big)$
where $F(r)$ is a dimensionless function parametrizing the skyrmion profile and $\vec{\hat{r}}$ represents the unit vector. Furthermore,
$\chi$ is assumed to take the form, $\chi(\vec{x})=f_\phi C(r)$, where $C(r)$ is also a dimensionless function.
For the soliton energy to be finite as $r\to \infty$ and for the ansatz to remain regular at $r=0$, the following boundary conditions are imposed, 
$F(r=\infty)=0$, $F(0)=\pi$, $C(r=\infty )=\chi_0/f_\phi$  and $dC(0)/dr=0$,
where $\chi_0$ is determined by the stationary condition of $\chi$ at $r=\infty$.
By numerically solving the coupled differential equations for $F(r)$ and $C(r)$, the skyrmion solution tagged with the baryon number of $1$ can be obtained.

According to Ref.~\cite{Carson:1991fu,GarciaMartin-Caro:2023klo}, 
substituting the soliton solutions into the EMT in Eqs.~(\ref{barEMTsChPT}) and (\ref{hatEMTsChPT}) yields the static EMT of the nucleon within the skyrmion approach, 
\begin{align}
    &\Theta^{\rm static}_{\mu\nu}(r)=
    \bar\Theta^{\rm Skyr}_{\mu\nu}(r)+\hat\Theta^{\rm Skyr}_{\mu\nu}(r)
    -\epsilon_{\mu\nu}^{\rm const}, \label{largeNc}
\end{align}
where $\bar\Theta^{\rm Skyr}_{\mu\nu}(r)$ and $\hat\Theta^{\rm Skyr}_{\mu\nu}(r)$ represent the traceless and trace part of EMT, obtained by inserting the soliton solutions. Note that the contribution of the Skyrme term is included in $\bar\Theta^{\rm Skyr}_{\mu\nu}(r)$. In addition, to ensure that the nucleon energy and pressure vanish as $r\rightarrow\infty$, we introduce $\epsilon_{\mu\nu}^{\rm const}$ in Eq.~(\ref{largeNc})
to eliminate constant contributions from the static EMT. 

For numerical calculations, this study uses the following model parameters.
The decay constant of pion and the Skyrme parameter are taken as $f_\pi=68 \ {\rm MeV}, \ e=5.45$ from Ref.~\cite{Adkins:1983ya}, which are determined by reproducing the experimental values of the nucleon mass and the delta mass within the skyrmion approach based on the ChPT.
The pion mass is fixed to be the experimental value $m_\pi=140 \ {\rm MeV}$.
The decay constant and mass of the scalar meson are taken from Ref.~\cite{Song:1997kx,Lee:2003eg}, with $f_\phi=240 \ {\rm MeV}, \ m_{\phi0}=720 \ {\rm MeV}$.
The value of $f_\phi$ is empirically taken to well describe the skyrmions within the sChPT framework and the value of $m_\phi$ is chosen to correspond to the lightest scalar meson.
For simplicity, this study ignores the anomalous dimension of the quark mass term, i.e., $\gamma_m=0$.
Even if $\gamma_m$ takes nonzero values, our main results remain unchanged. 
\footnote{
The results of other parameter sets, where the scalar meson mass $m_{\phi0}$ is varied, is presented in our full paper~\cite{Tanaka:2025pny}. Note that even with different parameter sets for the dilaton mass, the nucleon mass remains qualitatively unchanged.  
As an additional note, the parameter set used in our current analysis is in good agreement with a lattice observation of the GFF with momentum transfer, $D(t)$. Given this, we use the current model parameter set to evaluate the pressure and the D-term $D(0)$.
} 

\begin{figure}
\begin{center}
  \includegraphics[width=7.0cm]{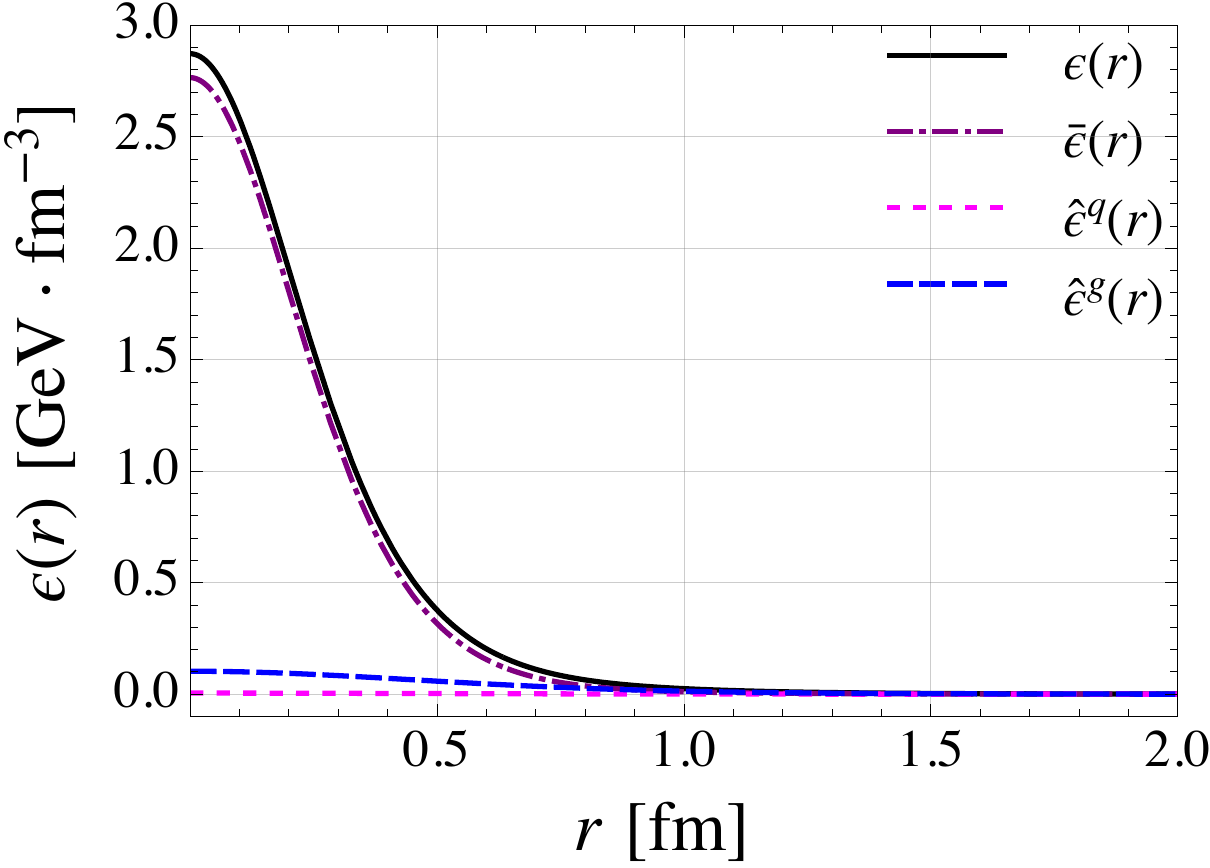}
    \subfigure{ (a) }    \includegraphics[width=7.0cm]{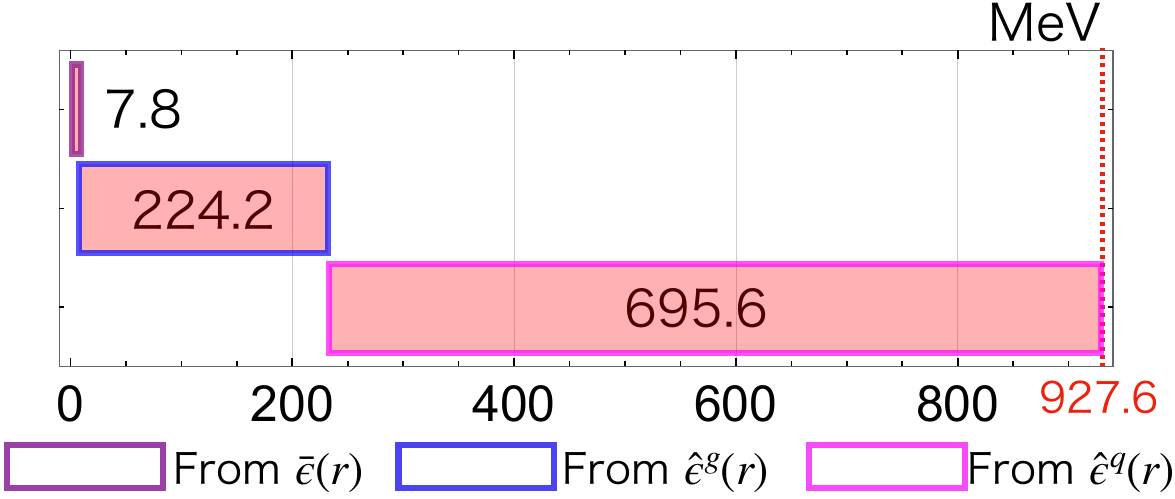}
    \subfigure{ (b) }
\end{center}
    \caption{
    (a) Spatial distribution of the energy density and its composition. (b) Nucleon mass composition. 
    }
    \label{energy_decomp}
\end{figure}
\section{Dominance of gluonic scale anomaly in confining pressure and D-term}

Before presenting the numerical results for the pressure and the D-term, 
we first show the spatial distribution of the energy density inside the nucleon and the nucleon mass in Fig.\ref{energy_decomp}, obtained using the skyrmion approach based on the sChPT.
Panel~(a) shows the 
spatial distribution of the energy density and its composition.
Compared to the scale anomaly contributions $\hat \epsilon^{q,g}$,
the dynamical energy part $\bar \epsilon$ is dominantly distributed. 
Performing the spatial integration of the energy distribution, the skyrmion mass is evaluated as $M_N=927.6 \ {\rm MeV}$. 
As shown in panel~(b), the nucleon mass composition is dominated by the dynamical energy mass, which accounts for $75\%$. This proportion is consistent with the analytical estimation in Eq.~(\ref{1/4M}).
Regarding the contribution from the scale anomaly, the gluonic contribution is the most significant, accounting for $24.2\%$.

\begin{figure}
\begin{center}
    \includegraphics[width=7cm]{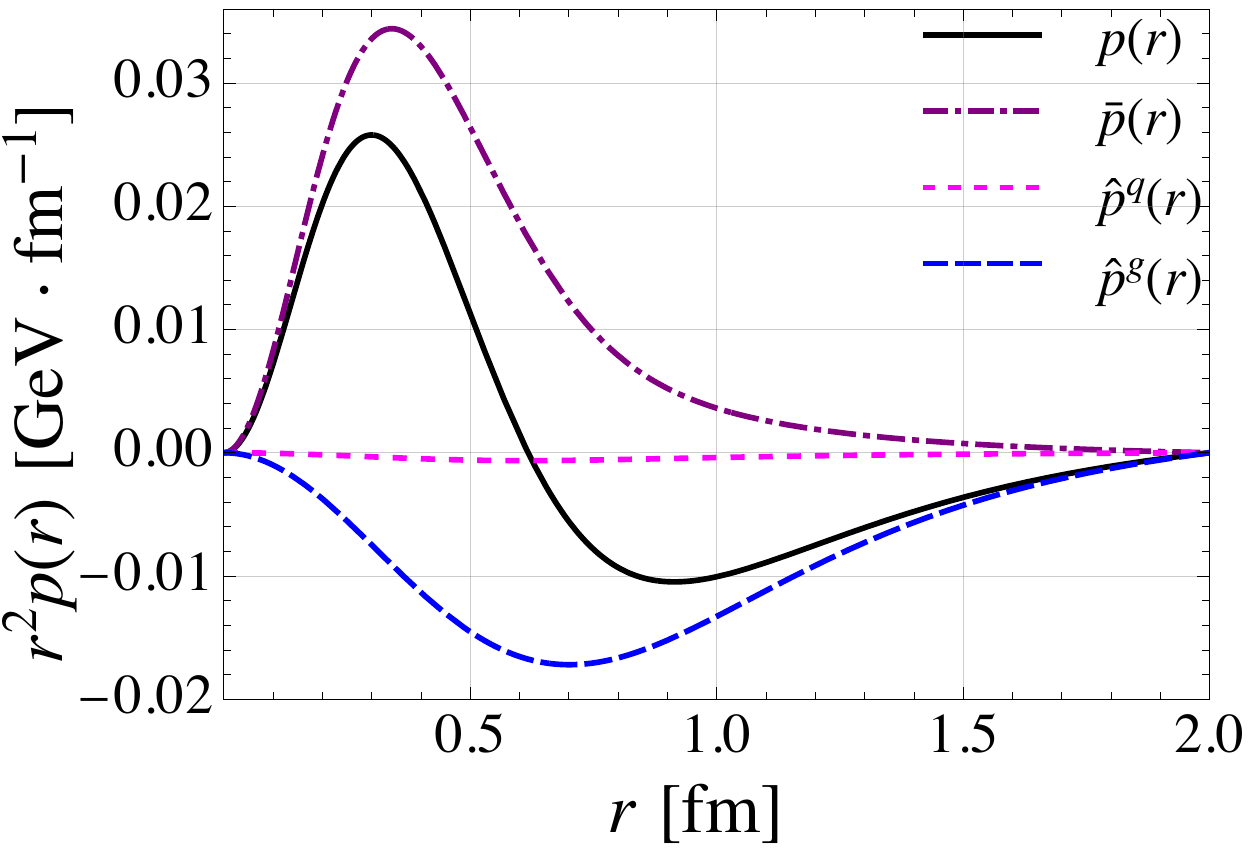}
    \subfigure{ (a) }
    \includegraphics[width=7cm]{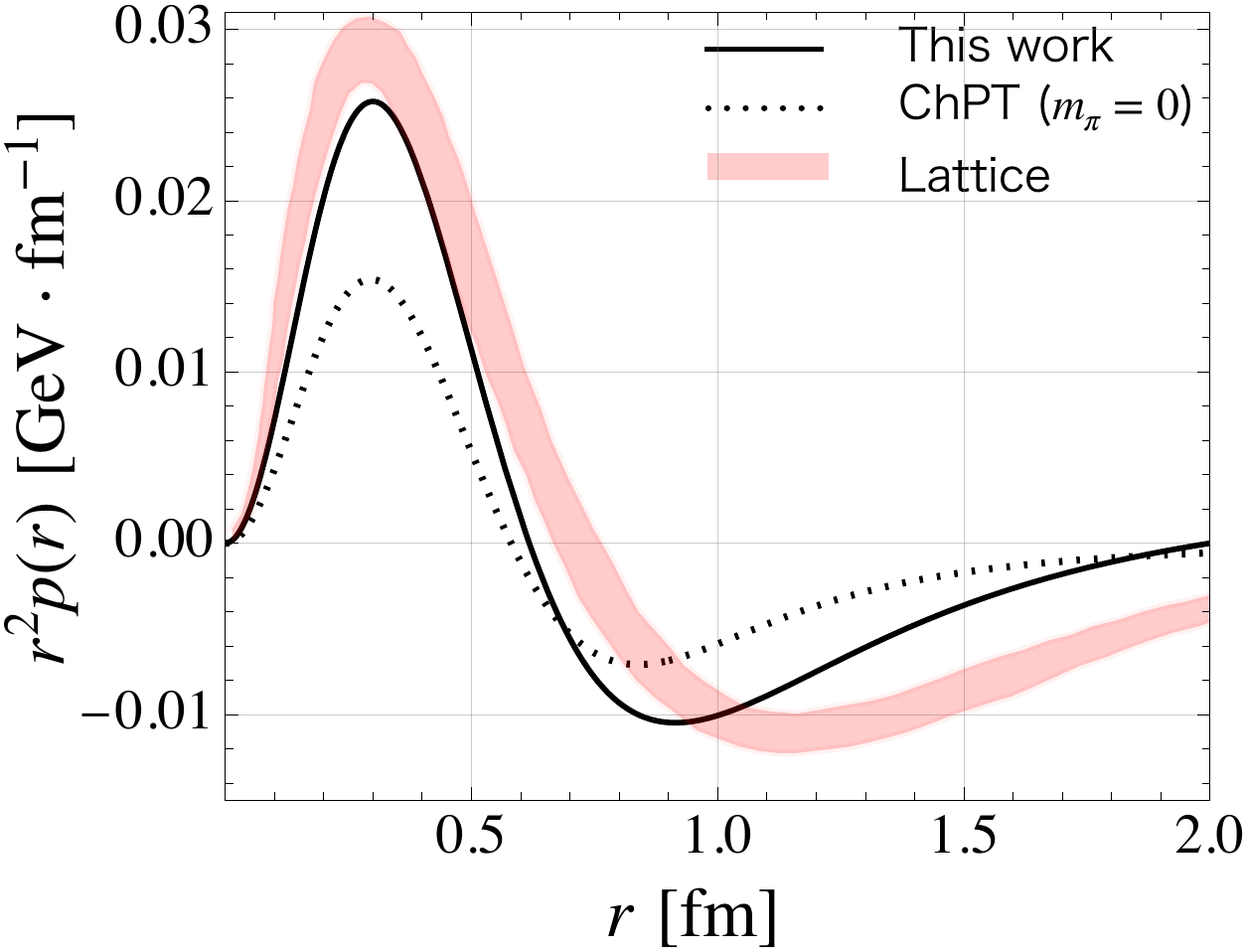}
    \subfigure{ (b) }
\end{center}
    \caption{
    (a)~Spatial distribution of the pressure and its composition. (b) Comparison with the lattice QCD observation~\cite{Shanahan:2018nnv}. 
    }
    \label{pressure_decomp}
\end{figure}

We then present the spatial distribution of the pressure $p(r)$ inside the nucleon and its composition in panel~(a) of Fig.~\ref{pressure_decomp}.
The dynamical energy contribution $\bar{p}(r)$ takes positive values throughout all spatial regions, acting locally as the repulsive force. In contrast, the scale anomaly contribution $\hat p^{q,g}$ takes negative values, leading to a role in confining.
As for this confining effect, the gluonic contribution is dominant and widely distributed.
In particular,
compared to other components, the pressure from the gluonic scale anomaly retains substantial values even at locations far from $r=0$.
At such distances, the confining force exceeds the repulsive force, causing the total pressure to turn negative as $r$ increases.

Having the effective model results, we compare the total pressure with the lattice observation~\cite{Shanahan:2018nnv} in panel~(b) of Fig.~\ref{pressure_decomp}.
To better elucidate the scale anomaly contributions, we also include the result from the skyrmion approach based on the ChPT in the chiral limit ($m_\pi=0$).\footnote{
In ChPT, the scale anomaly is not properly captured. ChPT with massive pion fails to reproduce even the current quark mass contribution to the scale anomaly in Eq.~(\ref{traceanomaly}). Moreover, the kinetic term $\mathcal{L}^{[2]}_{\mathcal{O}\left(p^2\right)}$ accidentally breaks the scale invariance.}
This figure shows that the ChPT result deviates from the lattice observations, whereas incorporating the scale anomaly within the sChPT framework improves the total pressure. 
The sChPT is capable of aligning more closely 
with the lattice QCD observation.

Given that our pressure is qualitatively aligned with the lattice QCD observation,
we proceed to estimate the D-term,
\begin{align}
    &D=-4.12. \label{Dvalue} 
\end{align}
This evaluation is consistent with the recent lattice QCD observation, within its systematic uncertainties, based on dipole fitting~\cite{Hackett:2023rif}. They also performed a z-expansion fitting, and our prediction is slightly smaller than their result based on that method. Additionally, recent model-independent calculations based on dispersion relations have also determined the D-term~\cite{Cao:2024zlf}, and our value is again slightly smaller. One possible reason for this discrepancy is our choice of $\gamma_m = 0$. We confirmed that when $\gamma_m > 0$, the D-term increases. A comprehensive discussion of this point is given in our full paper~\cite{Tanaka:2025pny}.

\begin{figure}
    \includegraphics[scale=0.4]{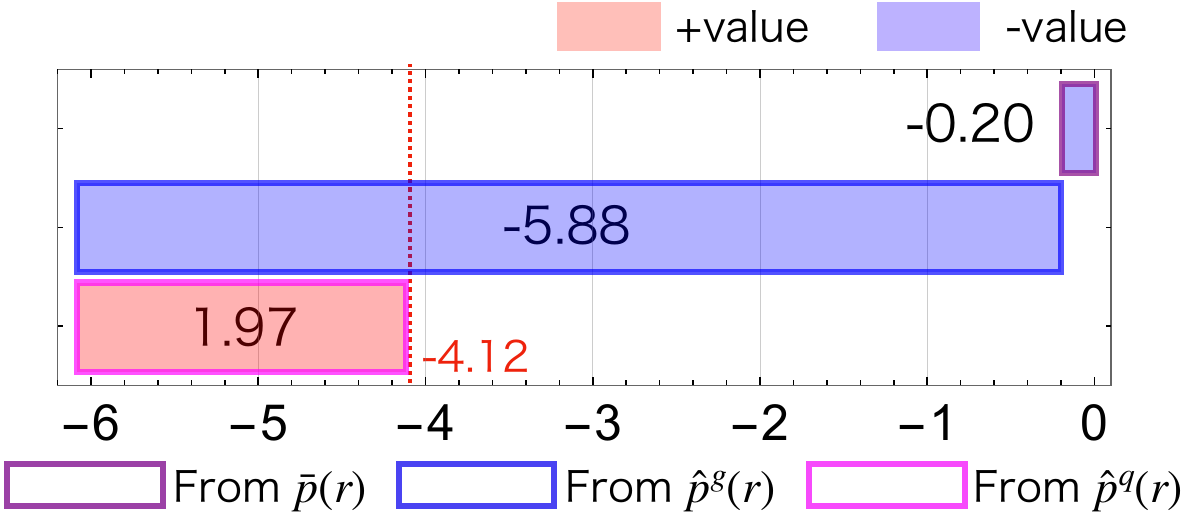}
    \caption{
    Composition of D-term
    }
    \label{Dterm_results}
\end{figure}

The composition of the D-term is shown in Fig.~\ref{Dterm_results}. 
Since $\hat p^{g}$ is widely distributed, as seen in panel~(a) of Fig.~\ref{pressure_decomp}, 
the mean square radius of $\hat p^g$ is relatively large, $\braket{r^2}_s^g=(0.52 \ {\rm fm})^2$, compared to the tensor mean square radius $\braket{r^2}_t=(0.30 \ {\rm fm})^2$.
In addition, the mean square radius of $\hat p^q$ is negligibly small, $\braket{r^2}_s^q=(0.097 \ {\rm fm})^2$. Hence, the negative D-term is driven by the contribution from the gluonic scale anomaly. From this result, it can also be understood that the D-term serves as a physical quantity that captures the spatial extents of the decomposed pressures.

\section{Conclusion}

In this letter, we analyzed the role of the scale anomaly in the stress distribution of the nucleon, particularly the pressure and the related D-term, using the skyrmion approach based on sChPT. As was mentioned in the introduction,  the gluonic scale anomaly has been found to dominantly contribute to the confining pressure inside the nucleon and the D-term. 
In the following, we provide a few comments related to our investigation. 

While the  stress distributions have already been explored for
various hadrons~\cite{Polyakov:1999gs,Brommel:2005ee,Broniowski:2008hx,Abidin:2008ku,Abidin:2008hn,Abidin:2009hr,Frederico:2009fk,Masjuan:2012sk,Yang:2014xsa,Son:2014sna,Fanelli:2016aqc,Hudson:2017xug,Shanahan:2018pib,Freese:2019bhb,Krutov:2020ewr,Shuryak:2020ktq,Kim:2020nug,Kim:2020lrs,Loffler:2021afv,deTeramond:2021lxc,Raya:2021zrz,Pefkou:2021fni,Krutov:2022zgg,Kim:2022wkc,Alharazin:2022wjj,Won:2022cyy,Hackett:2023nkr,Xu:2023izo,Li:2023izn,Allahverdiyeva:2023fhn,Czarnecki:2023yqd,Fujii:2024rqd,Krutov:2024adh,Broniowski:2024oyk,Liu:2024jno,Liu:2024vkj,Mamedov:2024tth,Hatta:2025ryj}, the analysis of the scale anomaly contribution based on the decomposition in Eq.~(\ref{staticEMTdecomp}) has not been conducted. 
Therefore, using lattice QCD and effective models~\cite{Park:2003sd,Park:2008zg,Haba:2010hu,Fujita:2022jus,Fujii:2024rqd} to investigate the extent to which the scale anomaly influences the confining pressure and the related D-term would be a valuable direction for further research, not only in the nucleon but also in other hadrons.

In panel~(b) of Fig.~\ref{pressure_decomp},
we compared the pressure result of sChPT with that of ChPT in the chiral limit 
by using our model parameter set. 
As shown in Ref.~\cite{Cebulla:2007ei,GarciaMartin-Caro:2023toa}, even when using a different parameter set for ChPT, the result remains nearly the same as our ChPT pressure. 
Considering the universal results in the ChPT, incorporating the scale anomaly contribution in ChPT is significant for aligning with the lattice QCD observation at short distances~$r$, while the sChPT results somewhat deviate from the lattice QCD observation at large distances.
It is worth noting that there remains room for discussion regarding the comparison with lattice QCD since
the lattice observations are obtained at a heavy pion mass~\cite{Shanahan:2018nnv}, which may lead to deviations from reality.
As an alternative to the pressure,
the recent lattice QCD observation near the physical pion mass has provided the longitudinal force density given by $p(r)+2s(r)/3$ with the shear force distribution $s(r)$
~\cite{Hackett:2023rif}. 
In fact, we also evaluated the longitudinal force density in the sChPT, which is in good agreement with the recent lattice QCD observation at larger distances, but not at shorter distances. This deviation may arise from $s(r)$. 
Indeed, there is a missing contribution
in the current sChPT model when evaluating the shear force distribution. In particular, the vector meson, especially the rho meson, has not been taken into account in the sChPT, yet it certainly contributes to the off-diagonal spatial components of the energy-momentum tensor.
Extending the model to include the vector meson based on the hidden local symmetry~\cite{Park:2003sd,Park:2008zg} is a valuable direction for future studies. A more detailed analysis of such extensions and comparisons is to be pursued in future work.
\\

Note added: 
After completing the present manuscript, we became aware of Ref.~\cite{Ji:2025gsq}, which has found that the gluonic scale anomaly dominantly contributes to the confining pressure. Remarkably, we independently reached the same result around the same time using the skyrmion approach. Furthermore, that work did not explicitly address the D-term.

\section*{Acknowledgments}

This work was supported in part by the Japan Society for the Promotion of Science (JSPS) KAKENHI (Grants No. JP24K17054) and the COREnet project of RCNP, Osaka University.
The work of M.K. is supported by RFIS-NSFC under Grant No. W2433019.

  \setcounter{section}{0}
  \setcounter{equation}{0}
  \setcounter{figure}{0}
  \renewcommand{\theequation}{A\arabic{equation}}
  \renewcommand{\thesection}{A\arabic{section}}
  \renewcommand{\thefigure}{A\arabic{figure}}


\bibliographystyle{apsrev4-1}
\bibliography{ref}

  \setcounter{section}{0}
  \setcounter{equation}{0}
  \setcounter{figure}{0}
  \renewcommand{\theequation}{S\arabic{equation}}
  \renewcommand{\thesection}{S\arabic{section}}
  \renewcommand{\thefigure}{S\arabic{figure}}



\end{document}